\documentclass[a4paper,12pt]{article}
\usepackage{epsfig}

\begin{document}

\def\beq{\begin{equation}}
\def\eeq{\end{equation}}
\def\bea{\begin{eqnarray}}
\def\eea{\end{eqnarray}}

\newcommand{\dedouble}{ \stackrel{ \leftrightarrow }{ \partial } }
\newcommand{\deR}{ \stackrel{ \rightarrow }{ \partial } }
\newcommand{\deL}{ \stackrel{ \leftarrow }{ \partial } }
\newcommand{\ci}{{\cal I}}
\newcommand{\ca}{{\cal A}}
\newcommand{\Wp}{W^{\prime}}
\newcommand{\vep}{\varepsilon}
\newcommand{\kk}{{\bf k}}
\newcommand{\pp}{{\bf p}}
\newcommand{\hs}{{\hat s}}
\newcommand{\proj}{\frac{1}{2}\;(\eta_{\mu\alpha}\eta_{\nu\beta}
+  \eta_{\mu\beta}\eta_{\nu\alpha} - \eta_{\mu\nu}\eta_{\alpha\beta})}
\newcommand{\projm}{\frac{1}{2}\;(\eta_{\mu\alpha}\eta_{\nu\beta}
+  \eta_{\mu\beta}\eta_{\nu\alpha}) 
- \frac{1}{3}\;\eta_{\mu\nu}\eta_{\alpha\beta}}

\newcommand{\PR}{P_{\mu\nu\alpha\beta}}
\newcommand{\trPR}{P_{\mu\mu\alpha\alpha}}
\newcommand{\tmumu}{T_{\mu\mu}}
\newcommand{\tmunu}{T_{\mu\nu}}
\newcommand{\GG}{G_{\mu\nu\alpha\beta}}
\newcommand{\GGm}{G^m_{\mu\nu\alpha\beta}}
\renewcommand{\thefootnote}{\fnsymbol{footnote}}
{\Large
\begin{center}
{\bf Rotationally non-invariant aspects  \\
of scattering amplitudes with graviton exchange}
% 11.30.-j  Symmetries and conservation laws
% 04.60.-m Quantum gravity
% 11.15.-q Gauge field theories
\end{center}}
\vspace{.3cm}

\begin{center}
Anindya Datta$^{1}$, $\;$ Emidio Gabrielli$^{2}$, $\;$
and $\;$ Barbara Mele$^{3}$ \\
\vspace{.7cm}

$^1$\emph{Harish-Chandra Research Institute,
Chhatnag Road, Jhunsi, \\
Allahabad, 211019, India}
\\
$^2$\emph{Helsinki Institute of Physics,
     POB 64, University of Helsinki, FIN 00014, Finland and
     CERN PH-TH, Geneva 23, Switzerland}
\\
$^3$\emph{Istituto Nazionale di Fisica Nucleare, Sezione di Roma,
and Dip. di Fisica, Universit\`a La Sapienza,
P.le A. Moro 2, I-00185 Rome, Italy}

\end{center}

\vspace{.3cm}
\hrule \vskip 0.3cm
\begin{center}
\small{\bf Abstract}\\[3mm]
\begin{minipage}[h]{14.0cm}
We consider the $s$-channel scattering 
of massive fermion or vector-boson pairs with
equal helicities,
mediated by a graviton in the linearized Einstein theory. 
We show that, although in general both  spin-2 and  spin-0
components are present in the exchanged graviton,
there is a special set of reference frames where
the spin-0 graviton 
  component vanishes. 
  This  is connected to the dependence of 
the trace of the graviton propagator on the effective dimension 
of the space-time spanned by the sources.
On the other hand,
one finds a non-vanishing (Lorentz-invariant)
interference of the graviton amplitude
with the scalar-exchange amplitude, that in principle could give measurable effects.
\end{minipage}
\end{center}
\vskip 0.3cm \hrule \vskip 0.5cm
\vskip 0.3cm
%%%%%%%%%%%%%%%%%%%%%%%%%%%%%%%%%%%
\section{Introduction}
%%%%%%%%%%%%%%%%%%%%%%%%%%%%%%%%%%%
In a previous work \cite{noi},
we considered  the {\it s-channel}
scattering amplitude of pairs of {\it on-shell}
massive fermions/gauge-bosons  mediated by a graviton,
in the linearized limit of quantum gravity in the 
Einstein theory. The amplitude is manifestly gauge invariant. 
We computed  its interference  with the corresponding amplitude where 
a scalar (Higgs boson) field replaces the graviton, and found a non-vanishing 
result.
This interference is of course a Lorentz-invariant quantity, 
that in principle could give measurable effects, and 
calls for  the presence of a spin-0 component
in the virtual graviton propagator.
\\
On the other hand,
a spin-0 virtual graviton component should correspond to 
 a non-vanishing trace in the {\it effective} graviton 
propagator. By {\it effective} we mean the ``active" part of 
the propagator, that is 
the one selected by the non-vanishing components of the source tensors.

In this paper,
we will show that, for the scattering process considered in Figure~1,
the spin content of the {\it effective} graviton propagator
 depends on the choice of the 
reference frame.
In particular, while in a {\it generic frame} 
one can trace back a spin-2 plus a spin-0
component in the effective propagator, there is a special
set of reference frames where the spin-0 component vanishes.
We will show explicitly this anomalous behavior by decomposing
the effective propagator in angular momentum eigenstates.
In particular, when projecting the angular momentum along the direction
orthogonal to the scattering plane in the c.m. frame, the spin-0
eigenstate is canceled by one of the spin-2 components.
The same holds also in any frame obtained from the latter by a Lorentz boost
orthogonal to the scattering plane, when projecting the spin along the boost
direction. 
\\
This cancellation resembles the one occurring in the {\it on-shell} limit
of the graviton propagator \cite{deser,VanNieuwenhuizen:1973fi}, that reduces
the spin degrees of freedom of the virtual graviton to the two $(\pm 2)$
helicities allowed for an on-shell (massless) graviton.

The appearance/disappearance of a spin-0 component, depending on the 
reference frame,
shows that rotational invariance  is not respected  by
the graviton-matter vertex, for  massive matter fields. This does not
contrast with what one expects from the Noether's theorem, since, as 
we will show, there is a
{\it negative-norm} (scalar) state contributing to the graviton propagator.
The latter evades the basic hypothesis of the quantum version of the 
Noether's theorem, where only positive-norm states are assumed. 

The plan of the paper is the following.
In Section~2, we study the dependence of the graviton propagator trace on the
effective dimensions of the source tensors, and notice the possibility of a {\it critical} dimension.
In Section~3, we give an explicit example of scattering processes where the source tensors can {\it live} in a critical-dimension subspace, giving rise to rotationally
non invariant aspects of the corresponding amplitude, as also shown in Section~4.
In Section~5, we present some concluding remarks.

For comparison, we discuss in Appendix A the corresponding photon exchange amplitude,
where, of course, no negative-norm state propagates and, consequently,
 rotational invariance holds. 
In Appendix B, we provide some relevant Lorentz covariant expressions for the matrix elements of the energy-momentum tensor.

%%%%%%%%%%%%%%%%%%%%%%%%%%%%%%%%%%%%%%%%%
\begin{figure}[ht]
\centerline{\epsfxsize=2.40truein \epsfbox{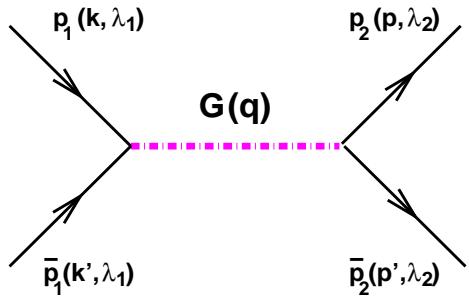}}
\caption{
\small
Graviton-mediated scattering of equal-helicity matter-field pairs.}
\label{uno}
\end{figure}
%%%%%%%%%%%%%%%%%%%%%%%%%%%%%%%%%%%%%%%%%%
%\vskip 1cm 
%%%%%%%%%%%%%%%%%%%%%%%%%%%%%%%%%%%%%%%%%%%%%%%%%%%%%%%%%%
\section{The trace of the graviton propagator}
%%%%%%%%%%%%%%%%%%%%%%%%%%%%%%%%%%%%%%%%%%%%%%%%%%%%%%%%%%
The  spin-0 component in the virtual graviton
can  be scrutinized by considering the trace of 
the graviton propagator tensor, that is naturally connected to the
rotationally invariant graviton component.
We will see that the effective propagator trace can vanish depending
on the {\it effective dimension} of the source tensors.

Let us consider the general form
 of the graviton propagator in the linearized
Einstein theory. When contracted with conserved energy-momentum tensors (i.e.,
for $q_{\mu}T_{\mu\nu}=0$, with $q_{\mu}$ the momentum flowing in the 
propagator), terms proportional to the momentum vanish,
 and the effective  graviton propagator 
becomes
\cite{veltman}
\beq
\GG(q)=i\, \;
\frac{\PR}
{q^2+i\epsilon} \; ; \;\;\;\;\;\;\; \;\PR=
\proj 
 \; ,
 \label{pro_G}
\eeq
where we use
the Minkowski metric $\eta_{\mu\nu}={\rm Diag}(1,-1,-1,-1)$.
\\
The $s$-channel  $p_{1}\bar p_{1}\rightarrow p_{2}\bar p_{2}$ scattering 
amplitude  mediated by a graviton
 is then given by
\beq
\ca=\frac{-i}{M_P^2}\left(\, {\hat T}^{\,f \dag}_{\mu \nu}\; 
\GG(q) \;{\hat T}^{i}_{\alpha \beta}\, \right) \;,
\label{AG}
\eeq
where ${ \hat T}^{\,i(f)}_{\mu\nu}$ are the matrix elements of the 
conserved energy-momentum
tensors between the vacuum and the initial (final) state, and 
$M_P$ is the Planck mass.

Apart from propagating spin-2 states,
the $\GG(k)$ tensor can  propagate a scalar  field  
associated to the trace  $\trPR$.

Let us now consider a given reference frame where the source tensor
${ \hat T}_{\mu\nu}$ has non-vanishing components
only in a $d$-dimensional subspace (with $d \le 4$).
Then, ${ \hat T}_{\mu\nu}$ can be decomposed as 
\beq
{\hat T}_{\mu \nu}={\hat R}_{\mu \nu} + 
{\hat T} \; \tilde{\eta}^{\,d}_{\mu\nu}\;,
\eeq
where $\;{\hat R}_{\mu \mu}=0\;$, $\;{\hat T}={\hat T}_{\mu \mu}/d\;$, 
and $\tilde{\eta}^{\,d}_{\mu\nu}$ is defined as follows: 
$\tilde{\eta}^{\,d}_{\mu\nu}={\eta}_{\mu\nu}\;$ for $(\mu,\nu)\,$ 
belonging to the
$d$-dimensional subspace, and $\tilde{\eta}^{\,d}_{\mu\nu}=0$ 
for all the other components. Note that $\tilde{\eta}^{\,d}_{\mu\nu}$
is not a tensor. Its definition depends on the reference frame.\\
Let $d^{\,\prime}$ (with $1 \le d^{\,\prime}\le 4$) be the 
smallest of the dimensions
relative to the two subspaces spanned by ${ \hat T}^{\,i}_{\mu\nu}$
and ${ \hat T}^{\,f}_{\mu\nu}$, respectively.
Then, let $d$ (with $d^{\,\prime}\le d \le 4$) be
the dimension of the minimal
subspace that contains  all the non-vanishing components of
both the matrix elements. Then,
the $\GG(k)$ tensor will  propagate a scalar  field  
associated to the trace
\beq
\tilde{\eta}^{\,d^{\,\prime}}_{\mu\nu} \; \PR \;\tilde{\eta}^{\,d}_{\alpha\beta} =
-\frac{1}{2}d^{\,\prime}\,(d-2)\;. 
 \label{traccia}
\eeq
The latter contributes
to the scattering amplitude in Eq.~(\ref{AG}) via
\beq
\ca^{J=0}=-\,\frac{(d-2)}{2 \, d \, M_P^2}\left(\,
\frac{\, {\hat T}^{\,f \dag}_{\mu\mu}\; {\hat T}^{i}_{\nu\nu}\,}
{q^2+i\epsilon} 
 \, \right)\; ,
\label{AGtraccia}
\eeq
whose coefficient depends on the reference frame, through the definition of
$d$. \\
The minus sign in Eq.~(\ref{AGtraccia}) is due to the negative norm of
the associated propagating (ghost) scalar field.

Let us now assume that there exists some reference frame where  the 
external sources {\it live} on the same plane, 
so that only their components
in a $d=2$ subspace can differ from zero
(we will give an explicit example of this case in Section~3).
Then, the scalar amplitude in Eq.~(\ref{AGtraccia}) vanishes.
In this case, no spin-0
graviton component will  propagate in the process.
\\
We observe that the absence of this spin-0 component does not occur
in any reference frame.
For instance, let us  rotate the above frame around 
an axis not orthogonal to the plane corresponding to the $d=2$ subspace,
where the source-tensor components can differ from zero.
Then, the source tensors will
acquire in general non-vanishing components along a third dimension. 
Correspondingly, the trace of the effective
propagator, according to Eq.~(\ref{traccia}), will be different from zero, 
giving rise
to the propagation of a spin-0 component of the graviton in 
Eq.~(\ref{AGtraccia}).

In Section~3, we will show that there are 
particular scattering processes, where the
source tensors can {\it live} in a two-dimensional
subspace, giving rise to the rotationally non-invariant aspects
 discussed above. We will show that the partial-wave decomposition of
 the amplitude includes a spin-0 component, and, at the same time,
in a particular frame,
the effective tensor $\PR$ propagating in the scattering
process is two dimensional and traceless, according to Eq.~(\ref{traccia}).
\\ Then, in Section~4,
we  will  analyze, in different frames, 
the rotational-group  representations contributing 
to the graviton polarization states that are effectively propagating 
in the process.

%%%%%%%%%%%%%%%%%%%%%%%%%%%%%%%%%%%%%%%%%%%%%%%%%%%%%%%%%%
\section{Scattering of equal-helicity matter fields}
%%%%%%%%%%%%%%%%%%%%%%%%%%%%%%%%%%%%%%%%%%%%%%%%%%%%%%%%%%

Let us consider, in its c.m. frame, the $s$-channel scattering, 
mediated by a graviton, of massive
fermion and  vector-boson pairs (Figure \ref{uno})
\beq
p_1\,(k,\lambda_1)\,+ \, \bar p_1\,( k^{\prime}, \lambda_1)\, \to
p_2\,(p,\lambda_2)\,+ \, \bar p_2\,( p^{\prime}, \lambda_2)\; ,
\label{fer}
\eeq
where initial(final) particles have equal and non-vanishing helicity  
$\lambda_1(\lambda_2)$. 
The 4-momenta of the initial and final  particles, $k/k^{\prime}$ and 
$p/p^{\prime}$, respectively, can be expressed
in terms of their spatial 3-momenta $\kk$ and $\pp$
\bea
k&=&(E,\kk),~~~ ~~~
k^{\prime}\;=\;(E,-\kk),
\nonumber \\
p&=&(E,\pp),~~~ ~~~
p^{\prime}\;=\;(E,-\pp),
\label{momenta}
\eea
with $E^2=|\kk|^2+m_1^2, \; \; E^2=|\pp|^2+m_2^2\;$, and $E=\sqrt{s}/{2}\;$.
\\
We then  define
the on-shell matrix element of the energy-momentum tensor
on an initial 
$p\bar p$ state, $\;\hat {\rm T}_{\mu\nu}[p,\bar p]\;$, as follows
\beq
{\rm \hat T}_{\mu\nu}[p,\bar p]\equiv  \;\; <0|T_{\mu\nu}|{ p, \bar p }> \; .
\label{sette}
\eeq
Due to the conservation of the 
on-shell matrix elements of the energy-momentum tensor, in the c.m. frame
one has ${\rm \hat T}_{0 \nu}=0$, 
and only the spatial components ${\rm \hat T}_{ij}\;\;\; (i,j=1,2,3)\;\;$
can differ from zero.

For equal-helicity fermions and {\it transverse} vector 
bosons\footnote{Only equal 
helicities for initial/final particles 
can give rise to $J=0$ components in the scattering amplitudes.}, 
the on-shell matrix elements of the energy-momentum tensor can be expressed
as  [see Appendix B]
\beq
{\rm \hat T}_{\mu\nu}[p,\bar p]\sim  \;\; (k-k^\prime)_{\mu} 
(k-k^\prime)_{\nu}\; ,
\label{kpk}
\eeq
(in case of {\it longitudinally} polarized vector bosons, 
${\rm \hat T}_{\mu\nu}[p,\bar p]\,$
presents an extra term proportional to $g_{\mu\nu}$).
The momentum $k(k^\prime)$ refers to the particle $p(\bar p)$.
In particular, in the c.m. frame,
for  fermion pairs of equal helicity $\lambda \;(=\pm 1)$, we find 
from Eqs.~(\ref{mat-uno})-(\ref{mat-uno-bis}) in the Appendix B
\beq
{\rm \hat T}_{0\nu}[f_\lambda,\bar f_\lambda]=0\;, \;\;\;\;\;\;
{\rm \hat T}_{i j}[f_\lambda,\bar f_\lambda]
=\left(\frac{4\lambda\, m_f }{\beta_f \sqrt{s} }\right) \, \kk_i\kk_j \; ,
%(i,j=1,2,3)
\label{EMT_f}
\eeq
where $\beta_f=\sqrt{1-m_f^2/E^2}$,  and $m_f$ is 
the fermion mass.

For  equal-helicity {\it transverse} real vector bosons, we obtain
from Eq.~(\ref{mat-due}) in the Appendix B
\beq
{\rm \hat T}_{0\nu}[V_{\pm}, V_{\pm}]=0\;, \;\;\;\;\;\;
{\rm \hat T}_{i j}[V_{\pm}, V_{\pm}]
=- \left(\frac{8\, m^2_V }{\beta_V^{\,2} \;s }\right) \, \kk_i\kk_j \; ,
%- 2\; m_V^2\;\; \frac{\kk_i\kk_j}{|\kk|^2} \; ,  
%(i,j=1,2,3)\; ,
\label{EMT_VT}
\eeq
where $\beta_V=\sqrt{1-m_V^2/E^2}$,  and $m_V$ is the vector boson mass.
\\
 Note that both the matrix elements 
 $\;{\rm \hat T}_{\mu \nu}[f_\lambda,\bar f_\lambda]\;$ and
 ${\rm \hat T}_{\mu \nu}[V_{\pm}, V_{\pm}]\;$ :
\begin{itemize}
\item {
vanish for massless states;}
\item {are proportional to $\kk_i\kk_j$, in the spatial sector.}
\end{itemize}
The amplitude for the gravitational scattering
 $p_{1}\bar p_{1}\rightarrow p_{2}\bar p_{2}$ is given by [cf. 
 Eq.~(\ref{AG})]
\beq
{\ca \Big[p_{1}\bar p_{1}\rightarrow p_{2}\bar p_{2}\Big]}
=-\frac{i}{M_{P}^2}\;\;
{\rm \hat T}^{\dag}_{\mu\nu}[p_{2}\bar p_{2}]\;\; G_{\mu\nu\alpha\beta}(q)\;\; 
{\rm \hat T}_{\alpha\beta}[p_{1}\bar p_{1}] \; ,
\label{amplitude}
\eeq
where $G_{\mu\nu\alpha\beta}(q)$ is the graviton propagator of momentum 
$q=k+k^{\prime}=p+p^{\prime}$.
By inserting  Eqs. (\ref{pro_G}), (\ref{EMT_f}) and (\ref{EMT_VT}) 
into Eq.(\ref{amplitude}),
we obtain the following expressions for the polarized scattering 
amplitudes in the c.m. frame
\bea
{\ca}\Big[f_\lambda \, \bar{f}_\lambda\to f^{\prime}_{\lambda^{\prime}}
\, \bar{f}^{\prime}_{\lambda^{\prime}}\Big]  &=&
    \frac{\hs}{6} \;\; \lambda\, \lambda^{\prime}  \beta_f\, \beta_{f^{\prime}}\, 
\; \sqrt{r_fr_{f^{\prime}}}
\;\; \left[  4\,{\bf d_{0,0}^{\,2}}(\theta) - {\bf d_{0,0}^{\,0}}(\theta)
\right] \; ,
\label{ffnoi} \\ 
%-------------
{\ca}\Big[f_\lambda \, \bar{f}_\lambda\to V_{\pm}\, 
 V_{\pm}\Big] 
&=& -\frac{\hs}{3} \;\;
 \lambda \;\beta_f\,\; \sqrt{r_f}\; r_V \;\; \left[4\,{\bf d_{0,0}^{\,2}}(\theta)
- {\bf d_{0,0}^{\,0}}(\theta)\right] \; ,
\\
%-------------
{\ca}\Big[V_{\pm} \,  V_{\pm}
\to V^{\prime}_{\pm}\,  V^{\prime}_{\pm}\Big] 
&=& \frac{2\,\hs}{3} \;\;
r_V r_{V^{\prime}}
\;\; \left[  4\,{\bf d_{0,0}^{\,2}}(\theta) - {\bf d_{0,0}^{\,0}}(\theta)
\right] \; ,
\label{jzero} 
\eea
where 
$${\bf d_{0,0}^{\,2}}(\theta)=\frac{1}{2}\left(3\cos^2{\theta}-1\right)\;\;\;\;\;\;\;\; 
{\rm and }\;\;\;\;\;\;\;\;\;\;
{\bf d_{0,0}^{\,0}}(\theta)=1
$$ 
are the Wigner functions (as defined in \cite{pdg})
corresponding  to the $J=2$ and $J=0$ 
eigenstates (with $J$ the total angular momentum), and
$\theta$ is the scattering angle between the final and
initial momenta, $\pp$ and $\kk$. Furthermore, we defined $\hs=s/M_{P}^2\;$,
$\;\; r_f=m_f^2/s$, and $r_{V^{(\prime)}}=m^2_{V^{(\prime)}}/s$.
$\;\; $
Eq.~(\ref{ffnoi}) has been previously derived in \cite{noi} 
for any fermion-helicity combination
(see, in particular, Eq.~(29) in \cite{noi}) .
%%%%%%%%%%%%%%%%%%%%%%%%%%%%%%%%%%%%%%%%%%%%%%%%%%%%%%%%%%
\section{Graviton polarization analysis in different frames}
%%%%%%%%%%%%%%%%%%%%%%%%%%%%%%%%%%%%%%%%%%%%%%%%%%%%%%%%%%
In this Section, we will show that, in case 
 the matrix elements of the energy-momentum
tensor in the amplitude in  Eq.(\ref{amplitude})
match the expression  in  Eq.(\ref{kpk}), 
the total angular-momentum decomposition of 
the  propagating  graviton  depends on the choice of the reference frame.
Note that rotational invariance  requires that 
the eigenstates of the $J^2$ Casimir operator
belonging to different $SO(3)$ representations do not mix under rotation.

In the c.m. frame, the scattering amplitude in  Eq.(\ref{amplitude})
is proportional to 
\beq
\ca \sim \sum_{i,j,\ell,m} {\rm \hat T}^{\dag}_{ij}(p)
\;\; P_{ij\ell m} \;\; {\rm \hat T}_{\ell m}(k) \, ,
\label{Ampkk}
\eeq
with
$$      
{\rm \hat T}^{\dag}_{ij}(p)\sim  \pp_i\, \pp_j \,,\;\;\;\;\;\;  \;\;
{\rm \hat T}_{\ell m}(k)\; \sim \; \kk_{\ell}\, \kk_m \,,
$$
where  the indices $(i,j,\ell,m)$ run from 1 to 3, 
$\;\kk$ and $\pp$ are defined as in Eq.~(\ref{momenta}), 
and the projector $P_{ij\ell m}$ is given by
\beq
P_{ij\ell m}=\frac{1}{2}(\delta_{im}\delta_{j\ell}+\delta_{i\ell}\delta_{jm}-
\delta_{ij}\delta_{\ell m}) \; .
\eeq
The fact that the spatial $T_{ij}$ matrix elements  are
proportional to 
the tensorial product of the corresponding external 
3-momenta, $\kk_i\kk_j$ and $\pp_i\pp_j$, makes 
the effective geometry of the problem {\it planar}.

In the following, the relevant 
angular-momentum representations for the graviton
polarization tensor in the 3-dimensional space, 
$\epsilon^{(J,J_3)}_{ij}$, are given by the 
$J=2$ and $J=0$ representations \cite{veltman,Spehler:1991yw}
\beq
\epsilon^{(2,+2)} =  \frac{1}{2}\left(
    \begin{array}{cl} 
      1~~~~i~~~0\\
     i~~-\!\!1~~0\\
     0~~~~0~~~0\\

\end{array}\right),~~~~~~~~
\epsilon^{(2,-2)} =  \frac{1}{2}\left(
    \begin{array}{cl} 
      1~~-i~~~~0\\
     -i~-1~~~0\\
     ~0~~~~~0~~~~0\\
\end{array}\right),
\eeq
%%%%%%%%%%%%%%%%%%%%%%%%%%%%%%%%%%%%%%%%%% 
\beq
\epsilon^{(2,\,0)} = - \sqrt{\frac{2}{3}}\left(
    \begin{array}{cl} 
     \frac{1}{2}~~~~0~~~0\\
     0~~~~\frac{1}{2}~~~0\\
     0~~~~0~-\!\!1\\
\end{array}\right),~~~~~~~~
\epsilon^{(0,\,0)} = - \frac{1}{\sqrt {3}}\left(
    \begin{array}{cl} 
     1~~~~0~~~0\\
     0~~~~1~~~0\\
     0~~~~0~~~1\\
\end{array}\right)\, ,
\label{jeq0}
%\\
\eeq
where $J_3$ is the  angular
momentum  projection along the 3rd axis.

If one 
chooses a frame where both $\kk$ and $\pp$ lie 
 in the $(1,2)$ plane,
the only non-vanishing components of the energy momentum tensors 
 are ${\rm \hat T}_{11},{\rm \hat T}_{22}$, and
${\rm \hat T}_{12}$. 
We can then write explicitly the sum in Eq.(\ref{Ampkk}) as
\beq
A\sim \frac{1}{2}\Big[{\rm \hat T}^{\dag}_{11}(p)-
{\rm \hat T}^{\dag}_{22}(p)\Big]\;
\Big[{\rm \hat T}_{11}(k)-{\rm \hat T}_{22}(k)\Big]+2
\,{\rm \hat T}^{\dag}_{12}(p)\, {\rm \hat T}_{12}(k) \; .
\label{Ampkk2}
\eeq
Hence, terms proportional to the trace
$[{\rm \hat T}^{\dag}_{11}(p)+
{\rm \hat T}^{\dag}_{22}(p)]\;[{\rm \hat T}_{11}(k)+
{\rm \hat T}_{22}(k)]$
do not contribute.
On the other hand, Eq.~(\ref{Ampkk2}) can be checked to be equivalent to the
 sum over the
virtual graviton polarizations 
\beq
A\sim \sum_{i,j,\ell,m} 
\Big\{\;{\rm \hat T}^{\dag}_{ij}(p)\, \Big[\, 
\epsilon_{ij}^{(2,+2)*}\,\epsilon_{\ell m}^{(2,+2)}\,
+ \epsilon_{ij}^{(2,-2)*}\,\epsilon_{\ell m}^{(2,-2)}\, \Big]\, 
{\rm \hat T}_{\ell m}(k)
\; \Big\} \; ,
\label{Ampkk3}
\eeq
where only the
$J=2, \;J_3=\pm 2$ representations of $SO(3)\;$ contribute in this case.

If, instead, one puts the initial particle 3-momenta along the 3rd axis,
Eq.(\ref{Ampkk}) becomes 
%\cite{wil}
\beq
A\sim -\frac{1}{2}\,
{\rm \hat T}^{\dag}_{33}(p)\;
\Big[{\rm \hat T}_{11}(k)-{\rm \hat T}_{33}(k)\Big]\; .
\label{Ampkk4}
\eeq
The decomposition of Eq.~(\ref{Ampkk4}) in terms of the graviton 
polarization tensors is then
\beq
A\sim \sum_{i,j,\ell,m} 
\Big\{\;{\rm \hat T}^{\dag}_{ij}(p)\, \Big[\, 
\epsilon_{ij}^{(2,\,0)*}\,\epsilon_{\ell m}^{(2,\,0)}\,
-\frac{1}{2}\, \epsilon_{ij}^{(0,\,0)*}\,\epsilon_{\ell m}^{(0,\,0)}\, \Big]\, 
{\rm \hat T}_{\ell m}(k)
\; \Big\} \; .
\label{Ampkk5}
\eeq
Note that, in Eq.~(\ref{Ampkk2}), one is projecting $J$ along the direction
orthogonal to the scattering plane, while, in Eq.~(\ref{Ampkk4}),
the $J_3$ direction in the one of the incoming particle momenta.

One can see that the presence of the $J=0$ representation 
in the effective graviton propagator depends
on the choice of the reference frame. This, of course, breaks 
the rotational invariance, but is not in contrast with the  Noether's theorem due to the presence of a negative-norm state.
This phenomenon is  connected in a straightforward way to 
the vanishing of the 
effective propagator trace in two dimensions, discussed in Section~2.
Indeed,  the second term in Eq.~(\ref{Ampkk5})
has a negative sign, arising from a ghost field. 

In Appendix A, for comparison, we discuss the corresponding $s$-channel amplitude
angular-momentum decomposition for the photon propagator. In the latter case,
no ghost field is present, and rotational invariance holds.
%%%%%%%%%%%%%%%%%%%%%%%%%%%%%%%%%%%%%%%%%%%%%%%%%%%%%%%%%%
\section{Concluding remarks}
%%%%%%%%%%%%%%%%%%%%%%%%%%%%%%%%%%%%%%%%%%%%%%%%%%%%%%%%%%
We have explicitly  shown that, in the $s$-channel amplitudes  
of massive fermion or vector-boson pairs with
equal helicities
mediated by a graviton, when projecting the graviton angular momentum
along the direction orthogonal to the scattering plane in the c.m. frame,
only the graviton polarizations corresponding to the $J_3=\pm 2$ projections
are exchanged [cf. Eq.~(\ref{Ampkk3})].  This angular momentum decomposition also holds in any frame
that is obtained from the latter by a 
Lorentz boost orthogonal to the scattering plane.
This decomposition  definitely contrasts with the presence of a $J=0$  component  both in Eq.~(\ref{Ampkk5}) and in  the partial-wave
decomposition of the  amplitudes in 
Eqs.~(\ref{ffnoi})-(\ref{jzero}). 
\\
On the other hand, in the Wigner functions entering
 Eqs.~(\ref{ffnoi})-(\ref{jzero}), the angular momentum is 
projected along the collision axis.
Then, the presence of a $J=0$ component 
in the partial-wave decomposition
seems to match the  Eq.~(\ref{Ampkk5}) content.
Nevertheless, in the partial-wave decomposition 
in Eqs.~(\ref{ffnoi})-(\ref{jzero}), 
the presence of a $J=0$ representation can not be rotated away
to recover Eq.~(\ref{Ampkk3}).
Indeed, the non-vanishing interference of the 
graviton  amplitude with the scalar exchange amplitude
computed in \cite{noi} [that is directly connected
to the projection of the graviton amplitudes 
in Eqs.~(\ref{ffnoi})-(\ref{jzero})
on the $J=0$ representation] 
is a Lorentz-invariant quantity.  The latter in principle could give measurable effects
in the Standard Model, where the scalar field is given by the Higgs boson \cite{noi}.

Note that  the decomposition  in Eq.~(\ref{Ampkk3}) 
can be viewed as the result of the cancellation 
in the ``active" two-dimensional (1,2) space
between
 the $J=0$ (ghost) component $\epsilon^{(0,\,0)}$ [arising in the graviton
 propagator in Eq.~(\ref{AGtraccia})]
and the $J=2, J_3=0$ polarization state corresponding to the 
$SO(3)$ representation  $\epsilon^{(2,\,0)}$ [cf. 
 Eq.~(\ref{jeq0})].

In conclusion, the Noether's theorem and the SO(3,1) invariance of 
the minimal
coupling of gravity in perturbation theory would imply the 
angular-momentum  
conservation  in the off-shell graviton coupling to matter fields \cite{landau},
in case off-shell (gauge-invariant)  ghost fields were absent.
\\
We found instead that the virtual graviton exchanged in the $s$-channel
scattering of {\it on-shell} massive matter fields
does not respect the  rotational invariance.
This effect is due to the presence, in the graviton propagator,
 of an off-shell (gauge-independent) scalar state
with negative norm, that evades the assumptions
required by the Noether's theorem.

%\newpage
%\vskip -0.5cm
\section*{Acknowledgments}
We acknowledge useful discussions with Lorenzo Cornalba.
We would like to thank  CERN PH-TH 
for its kind hospitality during the preparation of this work.
E.G. thanks Academy of Finland (Project No. 104368),
and CERN PH-TH for financial support.
This work was partially supported by the RTN European Programme
MRTN-CT-2004-503369 (Quest for Unification).
\vskip 2cm
\newpage
\section*{Appendix A}
As a comparative example, in this appendix we provide the angular momentum 
decomposition of the $s$-channel scattering amplitude $\ca^{J=1}$
mediated by a virtual photon. 
The corresponding Feynman diagram is the same as in 
Figure \ref{uno}, where the intermediate graviton propagator is
now replaced by the photon one. 
The corresponding amplitude is given by 
\bea
\ca^{J=1}=-i\,e^2\hat{J}_{\mu}^{f\dag}\, 
D^{\mu\nu}(q)\, \hat{J}^{i}_{\nu}\, ,
\label{AJ1}
\eea
where $e$ is the electromagnetic coupling, $q_{\mu}$ is the momentum
flowing in the photon  propagator $D_{\mu\nu}(q)$. The vectors
$\hat{J}^{\dag f}_{\mu}$ and $\hat{J}^{i}_{\mu}$ 
stand for the on-shell 
matrix elements of the electromagnetic current $J_{\mu}$
for the final and initial states, respectively.
The expression for $D_{\mu\nu}(q)$ in the $\xi$ covariant gauge is given by
\bea
D_{\mu\nu}(q)=\frac{i}{q^2+i\varepsilon}\left(-g_{\mu\nu} +(1-\xi)\, 
\frac{q_{\mu}q_{\nu}}{q^2}\right)\, .
\eea
Due to the conservation of the electromagnetic current 
($q^{\mu} \hat{J}^{\dag f}_{\mu}=q^{\mu} \hat{J}^{i}_{\mu}=0$),
the amplitude in Eq.(\ref{AJ1}) is gauge invariant,
and independent of the $\xi$ gauge-fixing term. Then, it 
can be simplified as follows
\bea
\ca^{J=1}=\frac{e^2}{q^2}\left(
-\hat{J}^{f\dag}_0 \hat{J}_0^i+ \hat{J}^{f\dag}_1 \hat{J}_1^i 
+\hat{J}^{f\dag}_2 \hat{J}_2^i+ 
\hat{J}^{f\dag}_3 \hat{J}_3^i\, .
\right)
\label{AJ1b}
\eea
Let us now consider a virtual photon with  momentum $q_{\mu}$  along 
the third axis, $q_{\mu}=(E,0,0,k)$.
By using the current conservation, implying
$E\, \hat{J}^{(f,i)}_0=k \,\hat{J}^{(f,i)}_3$, 
one can write Eq.(\ref{AJ1b}) as follows
\bea
\ca^{J=1}=\frac{e^2}{q^2}\left(
\hat{J}^{f\dag}_1 \hat{J}_1^i +\hat{J}^{f\dag}_2 \hat{J}_2^i + \frac{q^2}{E^2}
\hat{J}^{f\dag}_3 \hat{J}_3^i 
\right)\, .
\label{AJ1c}
\eea
The virtual photon contains both transverse and longitudinal polarizations.
The corresponding polarization vectors are given by
\bea
\varepsilon_{\lambda=+1}^{\mu}&=&\frac{1}{\sqrt{2}}\left(0,-1,-i,0\right)
\nonumber\\
\varepsilon_{\lambda=-1}^{\mu}&=&\frac{1}{\sqrt{2}}\left(0,1,-i,0\right)
\nonumber\\
\varepsilon_{\lambda=0}^{\mu}&=&\frac{1}{\sqrt{q^2}}\left(k,0,0,E\right)\, ,
\eea
where  $\lambda$ stands for the photon helicity, or 
the projection of the angular momentum $J$ along the third axis, $J_3$.
Then, by making use of the current conservation, the expression in 
Eq.(\ref{AJ1c}) can be rewritten as
\bea
\ca^{J=1}=\frac{e^2}{q^2}\,\hat{J}_{\mu}^{f \dag}
\left( \sum_{\lambda=-1}^{\lambda=1}  
\varepsilon_{\lambda}^{\star\mu}\varepsilon_{\lambda}^{\nu} \right)
\hat{J}_{\nu}^{i}\,,
\label{sum}
\eea 
where the sum  runs over the three $J=1$ polarization vectors.
Notice that for a real photon one has  $q^2=0$, and  only the 
$\hat{J}^{f\dag}_1 \hat{J}_1^i +\hat{J}^{f\dag}_2 \hat{J}_2^i$ 
contribution survives
in Eq.(\ref{AJ1c}),
corresponding to the two transverse physical polarizations of the photon.
This result 
 shows that only the components of the $J=1$ multiplet contribute
in the photon propagator. There is no $J=0$ component.

On the other hand, one can also look at  the partial wave 
decomposition of the 
photon exchange amplitude, and check that it
does not contain any $J=0$ component. As in  the graviton
case analyzed in this paper, let us consider the case of 
initial and final fermion pairs with equal 
helicities $\lambda (\lambda^{\prime})$ in the center of mass frame.
This helicity combination is the only one that can have a non vanishing
interference with a scalar-exchange amplitude. 
Then,  one has \cite{noi}
\bea
\ca^{J=1}\Big[f_{\lambda}\bar{f}_{\lambda}\to f^{\prime}_{\lambda^{\prime}}
\bar{f}^{\prime}_{\lambda^{\prime}}\Big]
%=4\sqrt{3}\,e^2\sqrt{r_f \,r_f^{\prime}}\, \beta_f\beta_{f^{\prime}}\, 
=\, \lambda \lambda^{\prime}\,e^2\,\sqrt{r_f \,r_f^{\prime}}\, 
\beta_f\beta_{f^{\prime}}\,\, 
{\bf d_{0,0}^{\,1}}(\theta) \, ,
\label{wave}
\eea
where  $r_{(f,f^{\prime})}=m_{(f,f^{\prime})}^2/q^2$, 
$\beta_{(f,f^{\prime})}=\sqrt{1-4m^2_{(f,f^{\prime})}/q^2}$, 
$\lambda^{(\prime)}=\pm 1$, and 
${\bf d_{0,0}^{\,1}}(\theta)=\cos{\theta}$ is the $J=1$ Wigner function 
as defined in
\cite{pdg}.
Hence, a scalar component is  present neither in Eq.(\ref{sum}) nor,
coherently,  in
Eq.(\ref{wave}).

There is no problem with  rotational invariance in the
photon exchange amplitude.
This is because, contrary to the graviton case,
in QED no negative-norm state is effectively propagating
in gauge-invariant amplitudes, and therefore
the requirements of the Noether's theorem regarding the angular momentum
conservation are satisfied.
\newpage

\section*{Appendix B}
In this Appendix, we provide the Lorentz covariant expressions for 
the matrix elements of the energy-momentum tensor ${T}_{\mu\nu}$ between 
the vacuum and a particle pair, as defined in Eq.~(\ref{sette}).
We consider two different  cases for the external particles:
fermions and  vector-bosons,  with non-vanishing masses
\cite{nnoi}.
\begin{itemize}
%-------------------------- FERMIONS ------------------------------
\item {\bf Fermions}
\bea
{\hat T}_{\mu \nu}[f_{\lambda}(k),\bar{f}_{\lambda^{\prime}}(k^{\prime})] &=&
\frac{1}{4}\left\{
{\hat J}^{\lambda\lambda^{\prime}}_{\mu}\, 
\left(k-k^{\prime}\right)_{\nu}+
{\hat J}^{\lambda\lambda^{\prime}}_{\nu}\, \left(k-k^{\prime}\right)_{\mu}
\right\}
\label{mat-uno}
\eea
where ${\hat J}_{\mu}^{\lambda\lambda^{\prime}}$ stands for
the matrix element of a $U(1)$ current 
$\bar{f_{\lambda^{\prime}}}(k^{\prime})\gamma_{\mu} f_{\lambda}(k)$
between the vacuum and a pair of fermion ($f_{\lambda}$)  and
anti-fermion ( $\bar{f}_{\lambda^{\prime}}$ ) of mass $m_f$ 
in the initial state,
with  momenta $k, k^{\prime}$ (entering the vertex), and helicities 
$\lambda, \lambda^{\prime}$, 
respectively.
In the c.m. frame, the vector 
${\hat J}_{\mu}^{\lambda\lambda^{\prime}}=
\bar{v}_{\lambda^{\prime}}(k^{\prime})\gamma_{\mu} u_{\lambda}(k)$
satisfies the condition ${\hat J}_{0}^{\lambda\lambda^{\prime}}=0$
due to the current conservation, and, up to a phase, the spatial components 
are given by
\bea
\hat{J}^{\lambda\lambda^{\prime}}_1&=&\,\delta_{\lambda,-\lambda^{\prime}}\,
 \sqrt{s}\, \Big(
\cos{\theta}\cos{\varphi}- i\, \lambda\sin{\varphi}\Big)\, 
 -2\lambda\, \delta_{\lambda,\lambda^{\prime}}\,m_f
\sin{\theta}\cos{\varphi}
\nonumber \\
\hat{J}^{\lambda\lambda^{\prime}}_2&=&\,\delta_{\lambda,-\lambda^{\prime}}\, 
\sqrt{s}\, \Big(
\cos{\theta}\sin{\varphi}+ i\, \lambda\cos{\varphi}\Big)\, 
-2\lambda\, \delta_{\lambda,\lambda^{\prime}}\, m_f
\sin{\theta}\sin{\varphi}
\nonumber \\
\hat{J}^{\lambda\lambda^{\prime}}_3&=&
 \,\delta_{\lambda,-\lambda^{\prime}}\, \sqrt{s}\, \Big(
-\sin{\theta}\sin{\varphi}\Big)\, 
-2\lambda \, \delta_{\lambda,\lambda^{\prime}}\, m_f 
\cos{\theta}
\label{mat-uno-bis}
\eea
where the 3-vector $\kk=-\kk^{\prime}$ is given by
$\kk=-\frac{\sqrt{s}}{2}\beta_f\Big(\sin{\theta}\cos{\varphi},\, 
\sin{\theta}\sin{\varphi},\,\cos{\theta}\Big)
$ and $\beta_f=\sqrt{1-4m_f^2/s}$\, .
%-------------------------- VECTORS ------------------------------
\item {\bf Real Vector Bosons}
\bea
{\hat T}_{\mu \nu}[V_{\vep}(k),{V}_{\vep^{\prime}}(k^{\prime})] &=&
\Big\{
\left(k\cdot k^{\prime}+{\rm m_V^2}\right)\left(\vep_{\mu}
\vep^{\prime}_{\nu}
-\frac{1}{2}\eta_{\mu\nu}\left(\vep\cdot \vep^{\prime}\right)\right)
\nonumber \\
&+&\frac{1}{2}\,\eta_{\mu\nu}\left(k\cdot \vep^{\prime}\right)
\left(k^{\prime}\cdot \vep\right)
-\vep^{\prime}_{\mu} k_{\nu} \left(k^{\prime}\cdot \vep\right)
\nonumber \\
&-&
\vep_{\mu} k^{\prime}_{\nu} \left(k\cdot \vep^{\prime}\right)
+\left(\vep\cdot \vep^{\prime}\right)\, 
k_{\mu} k^{\prime}_{\nu}
\Big\}
\,+\, \left\{\mu\leftrightarrow \nu\right\}
\label{mat-due}
\eea
where $\vep,\vep^{\prime}$ stand for  
vector-boson  polarization states, and $m_V$ is the vector-boson mass.
The  momenta $k, k^{\prime}$  are both entering the vertex.
\end{itemize}

\newpage
%%%%%%%%%%%%%%%%%%%%%%%%%%%%%%%%%%%%%%%%%%%%%%%%%%%%%%%%%%%%%%%%%%%%

\newpage

\end{document}